# Spatiotemporal mode-locking and photonic flywheel in multimode microresonators

Mingming Nie[2, *], Kunpeng Jia[1, *], Yijun Xie[2], Shi-ning Zhu[1], Zhenda Xie[1, *], Shu-Wei Huang[2, *]

[1]National Laboratory of Solid State Microstructures, School of Electronic Science and Engineering, College of Engineering and Applied Sciences, School of Physics, and Collaborative Innovation Center of Advanced Microstructures, Nanjing University, Nanjing 210093, China
[2]Department of Electrical, Computer and Energy Engineering, University of Colorado Boulder, Boulder, Colorado 80309, USA

[*]Corresponding author: mingming.nie@colorado.edu, jiakunpeng@nju.edu.cn, xiezhenda@nju.edu.cn, shuwei.huang@colorado.edu

Dissipative Kerr soliton (DKS) frequency combs – also known as microcombs – have arguably created a new field in cavity nonlinear photonics, with a strong cross-fertilization between theoretical, experimental, and technological research. Spatiotemporal mode-locking (STML) not only add new degrees of freedom to ultrafast laser technology, but also provide new insights for implementing analogue computers and heuristic optimizers with photonics. Here, we combine the principles of DKS and STML for the first time to demonstrate the STML DKS by developing an unexplored ultrahigh-quality-factor Fabry–Pérot microresonator based on graded index multimode fiber (GRIN-MMF). Using the intermodal stimulated Brillouin scattering, we can selectively excite either the eigenmode DKS or the STML DKS. Furthermore, we demonstrate an ultralow noise microcomb that enhances the photonic flywheel performance in both the fundamental comb linewidth and DKS timing jitter. The demonstrated fundamental comb linewidth of 400 mHz and DKS timing jitter of 500 attosecond (averaging times up to 25 μs) represent improvements of 25× and 2.5×, respectively, from the state-of-the-art. Our results show the potential of GRIN-MMF FP microresonators as an ideal testbed for high-dimensional nonlinear cavity dynamics and photonic flywheel with ultrahigh coherence and ultralow timing jitter.

Due to the low size, weight, power and cost and easy access to large comb spacing in nonconventional spectral ranges [1,2], Kerr microcomb has emerged as a promising frequency comb source and opens new applications such as highly multiplexed coherent optical communication [3,4], astrocombs [5,6], ranging [7,8], dual-comb spectroscopy [9,10], integrated frequency synthesizers [11,12], and optical clockwork [13,14]. Recently, a two-step pumping scheme for dissipative Kerr soliton (DKS) microcomb, utilizing the interplay between stimulated Brillouin laser (SBL) and cavity Kerr nonlinearity, is introduced to fundamentally narrow the comb linewidth and lower the repetition rate phase noise towards the quantum limit [15]. This two-step pumping scheme enables the photonic flywheel demonstration first in a monolithic fiber Fabry-Pérot (FP) cavity platform [15] and later in a silica disk microresonator [16] and a silica wedge microresonator [17].

Among all Kerr microcomb platforms, fiber FP platform provides a unique opportunity to study the spatiotemporal mode-locking enabled by the use of graded-index multimode fiber (GRIN-MMF). The parabolic core index profile of GRIN-MMF renders its transverse modes to cluster into nearly degenerate mode families whereas the intermodal dispersion between two mode families is orders of magnitude smaller than that of a regular step-index MMF or waveguide [18]. Such low modal dispersion enhances the intermodal nonlinear interaction and leads to the observation of new physical phenomena that attract growing interests including spatial beam self-cleaning and spatiotemporal mode-locking [19-24]. In particular, spatiotemporally mode-locked (STML) lasers not only add new degrees of freedom to applications such as telecommunications, imaging, and ranging [25], but also provide new insights for implementing analogue computers and heuristic optimizers with photonics [23].

Using an unexplored GRIN-MMF FP microresonator (Fig. 1a), we combine the principles of DKS with spatiotemporal mode-locking for the first time to demonstrate the STML DKS. We utilize the intermodal SBL to selectively excite either the eigenmode DKS or the STML DKS where DKSs of two different transverse modes are coherently locked to each other. Furthermore, we leverage the ultrahigh quality factor Q of $3.84\times10^8$ and use the two-step pumping scheme to achieve an ultralow noise Kerr microcomb that enhances the photonic flywheel performance [15] in both the fundamental comb linewidth, critical for optical atomic clockwork [13,14], and DKS timing jitter, benefiting microwave photonics and timing distribution [26,27]. The demonstrated fundamental comb linewidth of 400 mHz and DKS timing jitter of 500 attosecond (averaging times up to 25 μs) represent improvements of 25× and 2.5×, respectively, from the state-of-the-art [28,15]. Our results show the potential of GRIN-MMF FP microresonators as an ideal testbed for high-dimensional nonlinear cavity dynamics and photonic flywheel with ultrahigh coherence and ultralow timing jitter. Features and advantages of the GRIN-MMF FP microresonator platform are summarized in Supplementary Information Section I.

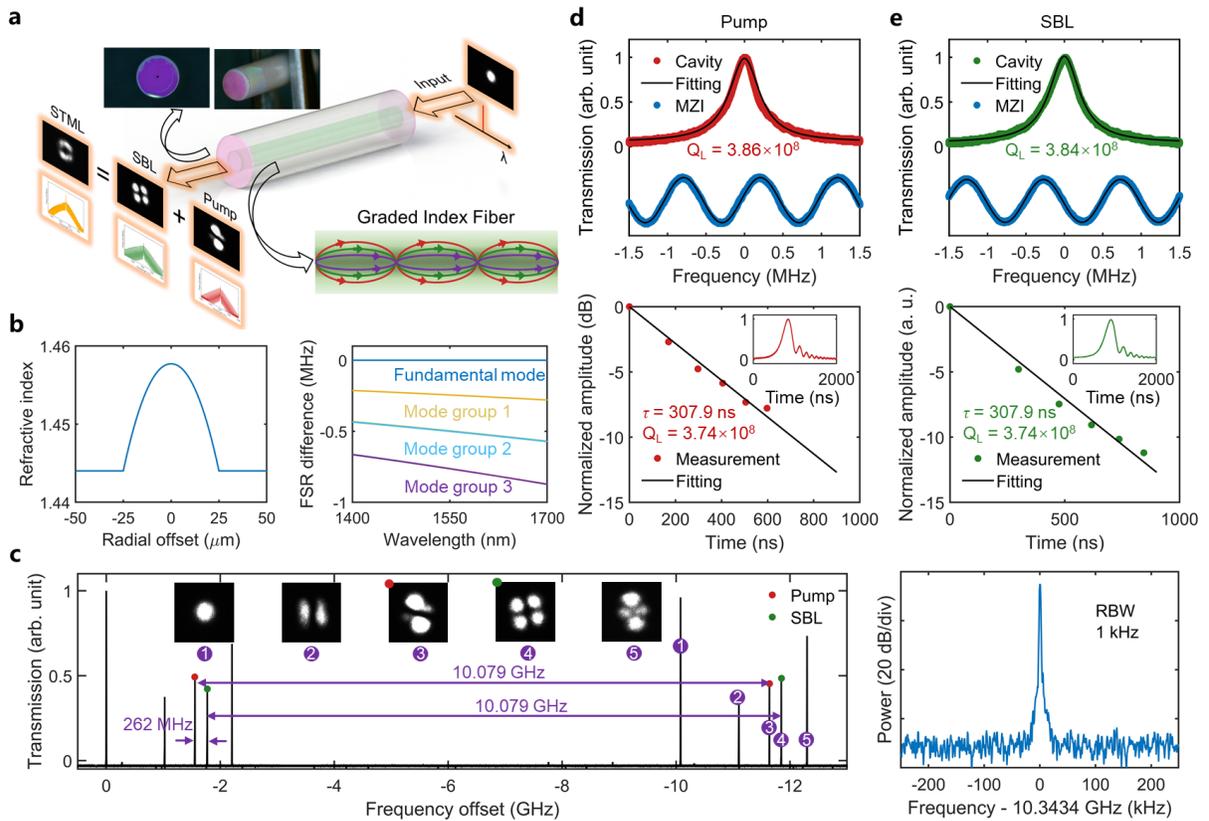

Fig. 1. Ultrahigh-Q GRIN-MMF FP microresonator. (a) Schematic architecture of the STML DKS. The inset at the upper left corner shows the photograph of the GRIN-MMF FP microresonator. The highly reflective dielectric Bragg mirror appears magenta. The inset at the lower right corner shows the propagation of three different transverse modes in the GRIN-MMF. (b) The left column shows the parabolic core index profile of the GRIN-MMF. The right column shows the cold cavity FSR difference between the three higher order mode families and the fundamental mode. (c) Left: transmission spectra of the five excitable transverse eigenmodes. Insets are the corresponding beam profiles measured with a CCD camera. FSR of 10.079 GHz at 1550 nm is measured. Right: SBS frequency shift of 10.343 GHz is measured. (d)(e) At ~1550 nm, the loaded Qs of both pump and SBL resonances are measured to be ~$3.8\times10^8$ by both the frequency-calibrated transmission spectra (top) and cavity ring-down traces (bottom).

Results
Ultrahigh-Q GRIN-MMF FP microresonators. Our ultrahigh-Q GRIN-MMF FP microresonator is fabricated through three steps: (i) commercial GRIN-MMF (GIF50E, Thorlabs) is carefully cleaved and encapsuled in a ceramic fiber ferrule; (ii) both fiber ends are mechanically polished to sub-wavelength smoothness [29]; (iii) both fiber ends are coated with optical dielectric Bragg mirror with reflectivity over 99.9% from 1530 to 1570 nm (Fig. 1a). Figure 1b shows the parabolic core index profile of the GRIN-MMF and the resulting small cold cavity free spectral range (FSR) difference between the first few linearly polarized (LP) mode families. By coupling a tunable external-cavity diode laser (ECDL, Toptica CTL1550) into the microresonator, we observe five specific transverse eigenmodes with good cavity loading as shown in Fig. 1c. The 20-mm long cavity length leads to a FSR of 10.079 GHz ± 2 MHz for all five recorded modes in Fig. 1c, ready for X-band microwave photonic applications.

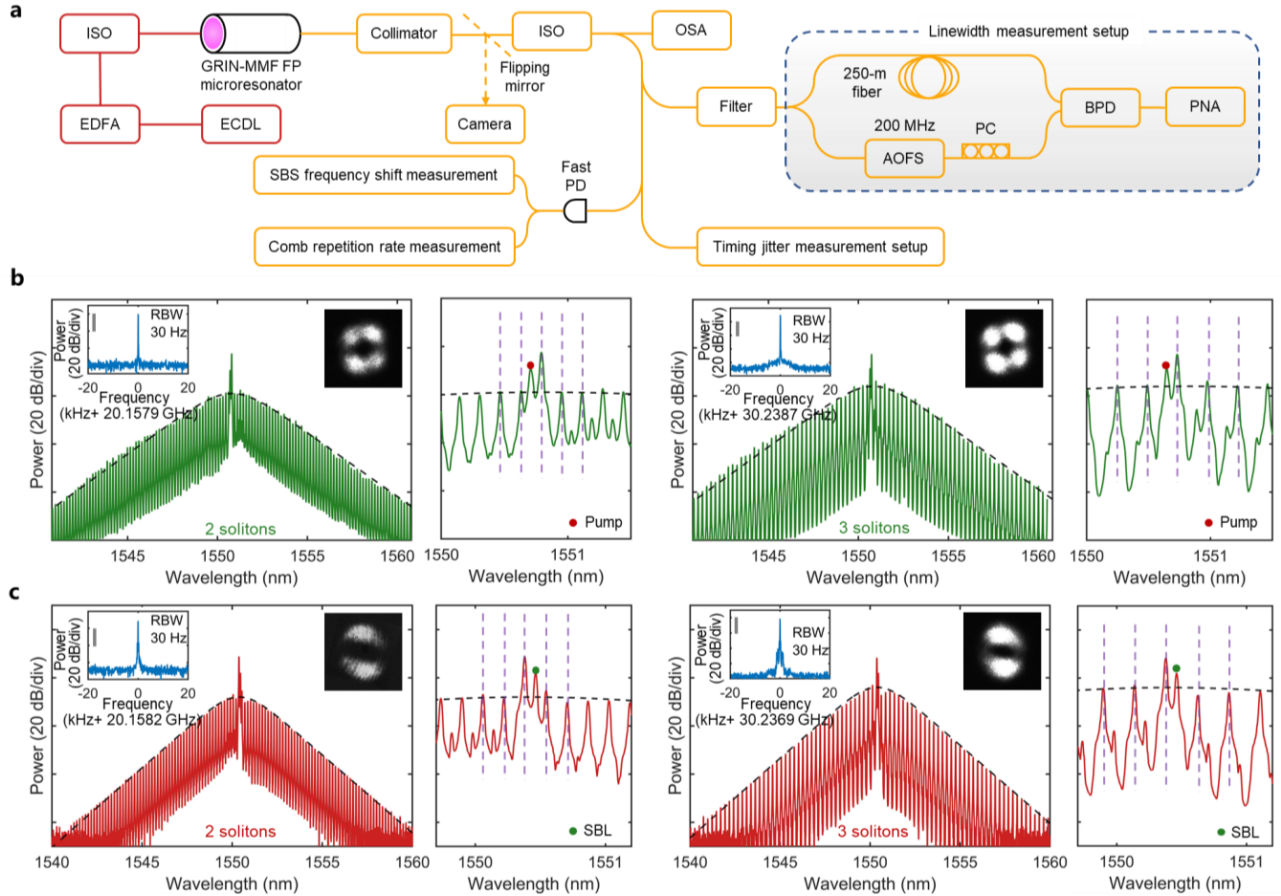

Fig. 2. Selective excitation of multimode DKS microcombs. (a) Schematic of the microcomb generation and characterization setup. ECDL: external-cavity diode laser, EDFA: Erbium-doped fiber amplifier, ISO: optical isolator, OSA: optical spectrum analyzer, PD: photodetector, AOFS: acousto-optic frequency shifter, PC: polarization controller, BPD: balanced photodetector, PNA: phase noise analyzer. (b)(c) Optical spectra of DKS crystals generated in the SBL mode family (b) and the pump mode family (c). The black dashed lines show the fitted DKS spectral envelope. The vertical purple dashed lines indicate the equidistant comb line positions. Insets are the RF beat note of comb repetition rate (left) indicating the stable mode-locking status and measured output beam profiles (right) indicating the selective excitation of different transverse mode families in Fig. 1c. Of note, comb lines induced by the cross-phase modulation (XPM) contain less than 1 % of the overall comb power.

Our previous FP microresonators were made of highly nonlinear fibers with small mode area, leading to high diffraction losses in the thick dielectric Bragg mirror coatings at both ends [15]. GRIN-MMF used in this work features a 20× larger mode area, and thus the diffraction loss is significantly mitigated to achieve ultrahigh Qs approaching $4\times10^8$ for all five modes (see Supplementary Information Section II). Figures 1d and 1e show both the frequency-calibrated transmission spectra and cavity ring-down traces [30] of the pump and SBL resonances, respectively. At ~1550 nm, the loaded Qs of both pump and SBL resonances are measured to be ~$3.8\times10^8$.

Selective excitation of multimode DKS microcombs. Figure 2a shows the microcomb generation and characterization setup schematic. Of note, all five transverse mode families exhibit similar anomalous GVD of -28 $fs^2$/mm according to the numerical simulation (see Supplementary Information Section II). Due to the ultrahigh Qs of all modes, the SBL threshold is measured to be as low as 15 mW.

To make the two-step pumping scheme work, the offset frequency between the pump and intermodal SBL should be conditioned slightly smaller than the SBS frequency shift such that the blue-detuned pump can compensate the thermal nonlinearity of the red-detuned comb generating SBL [15]. In our GRIN-MMF FP microresonator, such condition can be satisfied by using the 3$^{rd}$ and the 4$^{th}$ mode families for the pump and the SBL, respectively. In this configuration, the offset frequency (10.341 GHz, Fig. 1c) is 2 MHz smaller than the SBS frequency shift (10.343 GHz, Fig. 1c). Figure 2b shows the optical spectra, the beam profiles, and the radiofrequency (RF) beat note of the comb repetition rate when DKS crystals are generated in the SBL mode family. Single DKS generation (see Supplementary Information Section III) and comprehensive noise analysis will be discussed in the latter sections.

The two-step pumping scheme can also be generalized to facilitate the DKS comb generation in the pump mode family, as we demonstrate for the first time in Figure 2c. We fine-tune the cavity stress to increase the offset frequency between the pump and intermodal SBL and make it slightly larger than the SBS frequency shift. As discussed previously, all transverse mode families of the GRIN-MMF exhibit similar anomalous GVD and thus DKS microcomb can also be generated directly by the pump when it is put at the red detuning side of the resonance. In this configuration, the blue-detuned SBL is now used to compensate the thermal nonlinearity of the red-detuned and comb-generating pump. Using this cavity stress tuning method, we can achieve a selective excitation of DKS microcomb in an arbitrary transverse mode family.

The low modal dispersion and the inhomogeneous broadening of the SBS gain spectrum in the GRIN-MMF [31] greatly enhance the tolerance to the fabricated FSR variation and the flexibility in locating the proper pump-SBL pair that satisfies the requirement for the two-step pumping scheme. As shown in the Supplementary Information Section IV, we also achieve the selective DKS excitation in another ultrahigh-Q GRIN-MMF FP microresonator with a 10.087-GHz FSR, 8 MHz larger than the one in Fig. 2.

Spatiotemporally mode-locked DKS microcombs. Besides the selective excitation of multimode DKS microcombs, the low modal dispersion of GRIN-MMF FP microresonator provides another unique opportunity of generating STML DKS with low pulse energy and high repetition rates in the GHz regime. When the offset frequency between the pump and the intermodal SBL is set to be equal to the SBS frequency shift, the operation window within which both the pump and SBL are red-detuned and comb-generating are now open and accessible. As discussed previously, all transverse mode families of the GRIN-MMF exhibit similar FSR and thus STML DKS microcombs can be readily generated even without too much self-phase modulation (SPM) and intermodal XPM, facilitating the generation of STML DKS with low pulse energy and high repetition rates in the GHz regime.

Figures 3a and 3b shows the optical spectrum and the single comb-repetition-rate RF beat note of the first STML DKS microcomb ever demonstrated, respectively. The STML DKS microcomb can be categorized as the STML multimode soliton in MMFs [21,22], where multiple transverse modes are locked to share the same repetition rate but not the carrier-envelope offset frequency. This phenomenon is fundamentally different from the previous observations in STML fiber lasers [23,24], and it can be attributed to the microcomb's orders-of-magnitude higher FSR that consequently leads to orders-of-magnitude higher carrier-envelope offset frequency difference between transverse mode families. Thus, the intermodal nonlinear interaction in microcombs is not sufficient to compensate for the difference in the carrier-envelope offset frequency ($f_{ceo}$) but just the FSR. Such phenomenon is evidenced by the observation of the second RF beat note at ~262 MHz (Fig. 3c), corresponding to the carrier-envelope offset frequency difference ($\Delta f_{ceo}$) between the STML DKS microcombs in the pump and SBL transverse mode families. The single sideband (SSB) phase noise of the $\Delta f_{ceo}$ (Fig. 3d) confirms the phase coherence between the constituent transverse modes of the STML DKS, and the phase noise is currently limited by that of the pump laser itself.

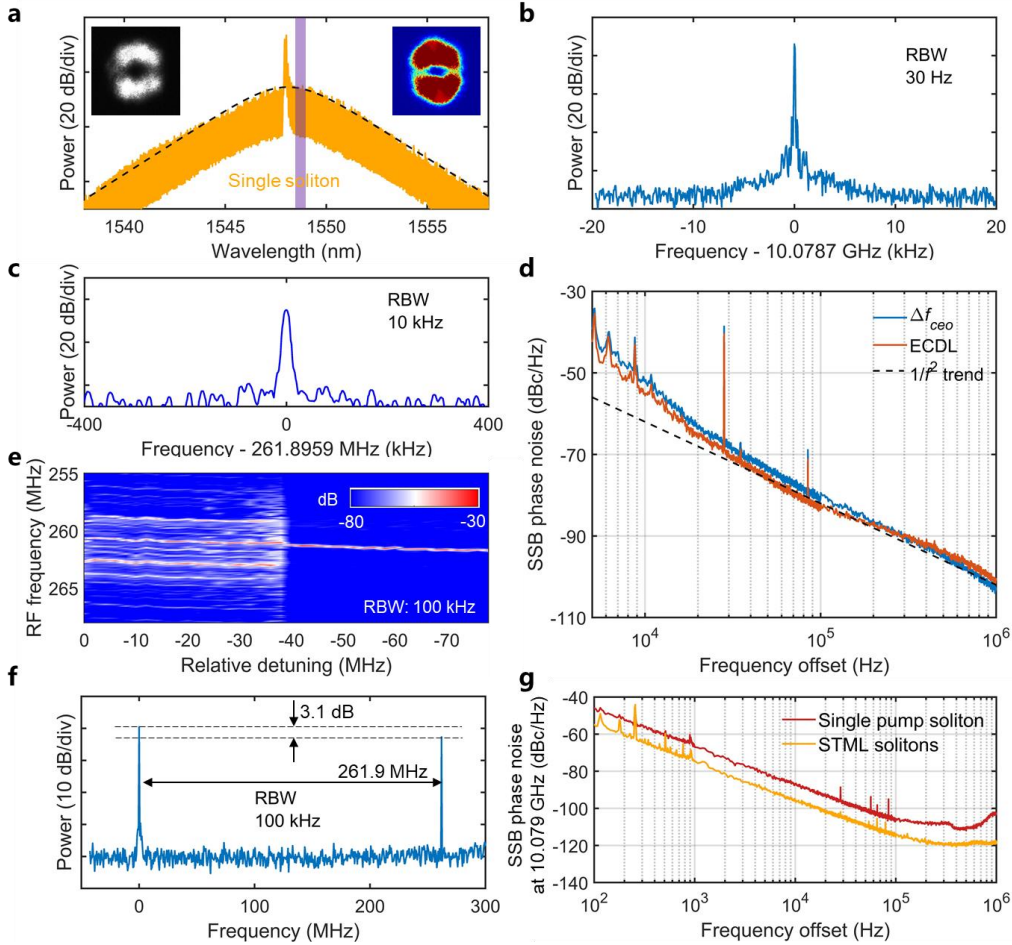

Fig. 3. Spatiotemporally mode-locked DKS microcombs. (a) Optical spectrum of the STML DKS. The black dashed lines show the fitted DKS spectral envelope. Insets are the measured (grey) and rebuilt (colored) beam profiles. The rebuilt beam profile is based on the superposition of the cold-cavity pump and SBL mode profiles in Fig. 1c with intensity difference of 3.1 dB. The purple shaded area indicates the optical bandpass filter. (b)(c) Clean and high-contrast RF beat notes of comb repetition rate and $\Delta f_{ceo}$, respectively. (d) SSB phase noise spectra of $\Delta f_{ceo}$ and ECDL. (e) Transition from the unlocked two-mode solitons (with multiple harmonics spaced by ~1.7 MHz) to STML DKS (with a single RF beat note). (f) RF beat notes between the comb lines and a tunable laser near 1548.3 nm, showing a 3.1-dB higher microcomb power in the SBL mode than the pump mode. (g) SSB phase noise spectra of the microcomb repetition rates, showing a 10-dB improvement of STML DKS over pump DKS.

We also depict the transition from the unlocked two-mode solitons to STML DKS by recording the electrical spectra around the $\Delta f_{ceo}$ while adjusting the pump-cavity detuning (Fig. 3e). Before the transition, several equally spaced RF beat notes around the $\Delta f_{ceo}$ are observed, characteristic of two independent microcombs with a repetition rate difference of ~1.7 MHz. When the pump-cavity detuning is further increased, a transition to STML DKS occurs and only one low-noise $\Delta f_{ceo}$ beat note is observed. The left and right insets in Fig. 3a show the beam profiles measured with a CCD camera and calculated by superposition of the 3rd and the 4th mode families, respectively. To determine the coefficients of the superposition, we beat the comb lines near 1548.3 nm with another ECDL to get the two RF peaks with frequency difference of 261.9 MHz and power difference of 3.1 dB (Fig. 3f).

In Fig. 3g, we characterize the repetition rate phase noises of pump DKS and STML DKS, which are limited to -110 dBc/Hz and -120 dBc/Hz at 1 MHz, respectively. We owe the 10 dBc/Hz improvement for the STML DKS to the synchronization induced phase noise reduction. Of note, all DKSs are initiated and self-stabilized over hours without any active control in the lab environment. Deterministic behavior is realized that the DKSs can be repeatedly and reliably generated following the same pump tuning protocol.

Photonic flywheel with a sub-hertz fundamental linewidth. Figure 4a compares the microcomb line SSB frequency noise spectra of pump and SBL DKSs, measured with an optical frequency discriminator using a fiber-based unbalanced Mach–Zehnder interferometer (UMZI) and a balanced photodetector (Fig. 2a, also see Methods and Supplementary Information Section V). Fundamental linewidths are then calculated from the white noise floor of the measured SSB frequency noise spectra. While the pump DKS simply inherits the frequency noise of the pump [32], yielding a fundamental linewidth of 130 Hz, both the SBL and the SBL DKS exhibit spectral purities that are more than two orders of magnitude better than the pump thanks to the ultrahigh Q of the GRIN-MMF FP microresonator. The 220-mHz SBL fundamental linewidth approaches that of the state-of-the-art on-chip SBLs (Table 1), but now at free running without any active stabilization of pump-cavity detuning.

Table 1. Comparison of SBL fundamental linewidth

| Material | Q factor (Million) | Linewidth (mHz) | Reference |
|---|---|---|---|
| $SiO_2$ (this work) | 384 | 220 | - |
| $SiO_2$ | 269 | 60 | 33 |
| $SiO_2$ | 140 | 376 | 34 |
| $SiO_2$ | 96 | 314 | 35 |
| $Si_3N_4$ | 29.2 | 720 | 36 |

Of note, the fundamental linewidths of the SBL DKS microcombs are slightly broadened from 220 mHz to ~400 mHz. This is because a larger red pump-cavity detuning is necessary for higher quality DKS, and the pump-cavity detuning induces imperfect phase matching in the SBS process that limits the linewidth narrowing factor [37]. Nevertheless, despite the compromise between the DKS microcomb linewidth and pulse quality, the 400-mHz fundamental linewidth demonstrated here still represents a 25× improvement over the state-of-the-art [16, 28, 38] (Table 2), benefiting applications including coherent optical communications and optical atomic clocks.

Photonic flywheel with a sub-femtosecond timing jitter. Figure 4b plots the SSB phase noise of the SBL DKS repetition rate, measured with the all-fiber reference-free Michelson interferometer (ARMI) setup providing attosecond timing jitter resolution [15,39,40] beyond the capability of direct photodetection methods [41-43] (see Methods and Supplementary Information Section VII). The measured SSB phase noises at 10 kHz, 100 kHz, and 1 MHz offset frequencies are -125 dBc/Hz, -148 dBc/Hz, and -168 dBc/Hz, respectively. The timing jitter integrated from 12 kHz to 1 MHz is 1 fs, which is only one fifth of the optical cycle at the DKS center wavelength. The DKS timing jitter approaches 500 attosecond on a 25-µs time scale that represents a 2.5× improvement over our previous photonic flywheel demonstration [15], benefiting applications including microwave photonics and timing distribution. Table 2 compares the repetition rate phase noise of the state-of-the-art microcombs. There is only one study in a $MgF_2$ crystalline microresonator [41] that shows a lower phase noise than ours. However, it is achieved using a transfer oscillator approach at the cost of expensive ultra-stable laser and complex active control electronics. Our approach, on the other hand, is completely free running without the need of any active control.

To find the dominant factor that limits the current DKS timing jitter, we simultaneously monitor the variations of SBL relative intensity noise (RIN), SBL frequency noise, SBL DKS RIN, and SBL DKS phase noise while we finely change the pump-cavity detunings (see Supplementary Information Section VI). SBL RIN is identified as the main limiting factor, and we project the SBL RIN transduced SSB phase noise (blue line) in Fig. 4b with the transduction coefficient calculated from the experimental parameters (see Supplementary Information Section VI). A good agreement is achieved with a maximum 6-dB deviation at offset frequencies below 1 kHz where other slow technical noises also start to contribute. The deviation for offset frequencies above

1 MHz is attributed to the pump-SBL interference and the coherent artefacts of the ARMI setup (see Supplementary Information Section VII). With an active control of the SBL RIN [35,47], we expect that a quantum-limited SSB phase noise can be achieved which will then be a 20-dB improvement over the state-of-the-art thanks to the large mode volume of the GRIN-MMF FP microresonator.

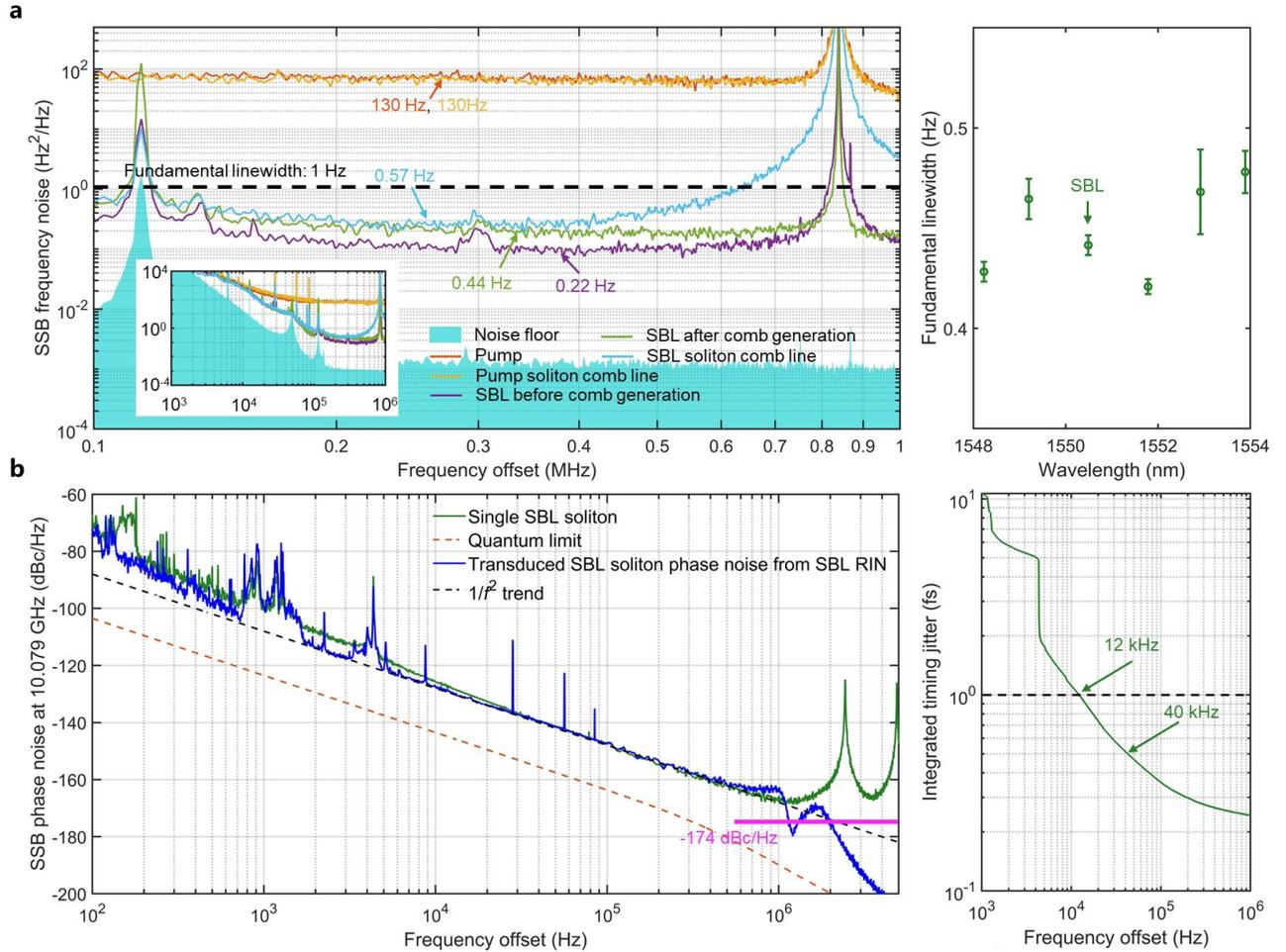

Fig. 4. Fundamental linewidth and timing jitter of the SBL DKS microcomb. (a) SSB frequency noise spectra of pump and SBL DKSs (left), and the wavelength dependence of the SBL DKS microcomb line fundamental linewidth (right). (b) SSB phase noise of the SBL DKS repetition rate, overlaid with the quantum noise limit and SBL-RIN-transduced repetition rate phase noise (left), and the integrated timing jitter with the dashed line representing the 1-fs timing jitter level (right). The ARMI setup provides attosecond timing jitter resolution with a −174 dBc/Hz phase noise floor at 10.079 GHz carrier frequency. The coherent artefacts at 2.5 MHz and its harmonics result from the 82-m delay fiber used in the ARMI setup.

Table 2. Comparison of DKS microcomb performances

| Fundamental linewidth | | | | |
|---|---|---|---|---|
| Material | Q factor (Million) | Microcomb linewidth (Hz) | Soliton active control | Reference |
| SiO$_2$ (this work) | 384 | ~0.4 | W/o | - |
| SiO$_2$ | 44.4 | 24 | W/o | 16 |
| Si$_3$N$_4$ | 56 | ~10 | W/o | 28 |
| Si$_3$N$_4$ | 11.6 | ~1000 | W/ | 38 |

| Phase noise of microcomb repetition rate | | | | | | |
|---|---|---|---|---|---|---|
| Material | Configuration | Carrier frequency (GHz) | SSB phase noise (dBc/Hz, scaled to 10 GHz) | | Soliton active control | Reference |
| | | | 10 kHz | 100 kHz | | |
| SiO$_2$ (this work) | Bright soliton | 10.08/20.16 | -125 | -148 | W/o | - |
| MgF$_2$ | Bright soliton | 14.09 | -142 | -159 | W/, (complex) | 40 |
| SiO$_2$ | Bright soliton | 0.945 | -120 | -140 | W/o | 15 |
| SiO$_2$ | Bright soliton | 10.43 | -125 | -144 | W/o | 16 |
| SiO$_2$ | Bright soliton | 11.02 | -120 | -139 | W/o | 17 |
| Si$_3$N$_4$ | Dark soliton | 5.4 | -108 | -134 | W/o | 28 |
| SiO$_2$ | Bright soliton | 22 | -111 | -147 | W/ | 40 |
| SiO$_2$ | Bright soliton | 15.2 | -117 | -143 | W/ | 44 |
| MgF$_2$ | Bright soliton | 9.9 | -130 | -130 | W/ | 45 |
| Si$_3$N$_4$ | Bright soliton | 9.78 | -110 | -130 | W/ | 46 |

Discussion

In summary, we develop a compelling DKS platform based on ultrahigh-Q GRIN-MMF FP microresonators. Utilizing the intermodal SBS, we selectively excite the SBL DKS, the pump DKS, and the first ever demonstrated STML DKS. Leveraging the ultrahigh Q and large mode volume of the GRIN-MMF FP microresonator and using the two-step pumping scheme, we achieve an ultralow noise microcomb that enhances the photonic flywheel performance in both the fundamental comb linewidth and DKS timing jitter. The demonstrated fundamental comb linewidth of 400 mHz and DKS timing jitter of 500 attosecond (averaging times up to 25 μs) represent improvements of 25× and 2.5×, respectively, from the state-of-the-art. With the new transverse mode degrees of freedom added to microcombs, the GRIN-MMF FP microresonator is a promising platform for applications such as space division multiplexing in telecommunications [48]. Our results show the potential of GRIN-MMF FP microresonators as an ideal testbed for high-dimensional nonlinear cavity dynamics and photonic flywheel with ultrahigh coherence and ultralow timing jitter.

Methods

Fundamental linewidth measurement We measure the frequency noise and fundamental linewidth of the soliton microcombs based on a self-heterodyne frequency discriminator using a fiber-based UMZI and a BPD. As shown in Fig. 2a, one arm of the UMZI is made of 250-m-long single mode fiber, while the other arm consists of an AOFS with frequency shift of 200 MHz and a polarization controller for high-voltage output. The FSR of the UMZI is 0.85 MHz. The two 50:50 outputs of the UMZI are connected to a BPD (Thorlabs PDB570C) with a bandwidth of 400 MHz to reduce the impact of detector intensity fluctuations. The detector output is then analyzed by a phase noise analyzer (RDL NTS-1000A). The relationship between SSB frequency noise PSD $L_\nu(f)$ in the unit of Hz$^2$/Hz and PNA output SSB phase noise PSD $L_\phi(f)$ in the unit of dBc/Hz, is given by [36,37]

$$L_\nu(f) = \frac{f^2}{4\sin^2(\pi f \tau)} L_\phi(f), \quad (1)$$

where $\tau$ is the delay time (~1.18 μs) of UMZI and $f$ is the frequency offset. The fundamental linewidth of the comb line $\Delta\nu$ in unit of Hz, is given by:

$$\Delta\nu = \pi L_{\nu w}, \quad (2)$$

where $L_{\nu w}$ is the value of frequency noise where $L_\nu(f)$ is flat, indicating white frequency noise.

We test our optical frequency discriminator with a known sub-hertz level reference laser at 1535 nm [49]. The measured result of 0.3 Hz (see Supplementary Information Section V) indicates the reliability of our linewidth measurement setup for sub-hertz fundamental linewidth. We also measure the noise floor of the setup with SBL before soliton generation by a balanced MZI with identical 250-m-long fiber spool for each arm. The noise floor $L_{\nu N}(f)$ is given by [50]

$$L_{\nu N}(f) = L_\phi(f)/4\pi^2\tau^2, \quad (3)$$

and is shown by the shaded area in Fig. 4a. Large contrast of >20 dB between the setup noise floor and the measured frequency noise, indicates the reliability of the sub-hertz linewidth results. In order to measure the linewidth of comb teeth, a tunable optical filter with 10 GHz bandwidth is employed to extract the desired comb tooth.

Measurement of comb repetition rate phase noise and timing jitter. The RF beat notes of comb repetition rate in Fig. 2 are realized by injecting the filtered comb lines into a RF-modulated electro-optic intensity modulator (EOIM), detecting by a fast PD (>10 GHz bandwidth) and analyzing the down-shifted beat note by an electrical spectrum analyzer. However, the EOIM will induce extra noise to the phase noise of the measured comb repetition rate. Therefore, for precise measurement of pump soliton and the STML soliton, we inject the comb lines into the fast PD, electrically divide the 10.079 GHz RF signal by 8 times (with a low noise floor) to 1.26 GHz and then analyze the divided signal with a PNA (see Supplementary Information Section VII).

Since the soliton phase noise measurement based on fast PDs is limited not only by the shot noise but also the available electronics operating at high frequency, we introduce ARMI setup for precise SBL soliton phase noise. Of note, the ARMI setup is not suitable for relatively high phase noise measurement, which invalidates the linear approximation during the process of converting phase noise to intensity noise. The details of ARMI setup can be found in the Supplementary Information Section VII.

Quantum-limited timing jitter. The quantum-limited SSB phase noise PSD of soliton repetition rate is given by [40,51]

$$L_{\phi QN}(f) = \frac{\sqrt{2}\pi}{2}\sqrt{\frac{\gamma}{\Delta_0(-D)}}\frac{g}{\eta\gamma^2}$$
$$\times\left[\frac{1}{96}\frac{\gamma(-D)}{\Delta_0}\frac{\eta\gamma^2}{f^2} + \frac{1}{24}\left(1+\frac{\pi^2 f^2}{\gamma^2}\right)^{-1}\frac{\eta\gamma^2}{\pi^2 f^2}\frac{\Delta_0(-D)}{\gamma}\right], \quad (4)$$

where $2\gamma$ ($=2\pi\times 0.5\times 10^6$ rad/s) is the FWHM resonance linewidth, $\Delta_0$ ($=2\pi\times 2\times 10^6$ rad/s) is the soliton detuning, $g$ ($=1.513\times 10^{-4}$ rad/s) is the frequency shift of a resonant mode per photon, $D$ ($=-0.0115$) is the normalized GVD of soliton mode, and $\eta$ ($=1$) is the quantum efficiency of the detector. As the ARMI system is not limited by the shot noise of PD, the shot-noise term is removed from Eq. 68 of Ref. [51]. According to Eq. (4), quantum noise can be reduced by smaller $g$ and $D$. Since $g = \bar{h}\omega_0^2 c n_2/V n_0^2$, where $\bar{h}$ is the Plank constant, $\omega_0$ is the DKS center frequency, $c$ is the speed of light, $n_2 = 3.6\times 10^{-20}$ m$^2$/W is the nonlinear Kerr parameter, $V = 5\times 10^{-12}$ m$^3$ is the mode volume and $n_0 = 1.4682$ is refractive index, the feasible way to lower the quantum limit is to generate soliton combs in a microresonator with small nonlinearity and large mode volume, which requires loss reduction and Q factors enhancement for low threshold.

## Data availability
All data generated or analyzed during this study are available within the paper and its Supplementary Information. Further source data will be made available on reasonable request.

## Code availability
The analysis codes will be available on reasonable request.

## Acknowledgments
We thank Professor S. A. Diddams from NIST, Boulder for fruitful discussions. We thank Professor I. Coddington and N. R. Newbury from NIST, Boulder for the reference laser at 1535 nm. We thank Dr. Dohyeon Kwon for helpful discussion regarding the ARMI setup. We thank Dr. L. G. Wright for fruitful discussion regarding the spatiotemporal mode-locking. We thank Dr. Bowen Li for fruitful discussion on the manuscript revision. M.N., Y.X., and S.W.H. acknowledge the support from the University of Colorado Boulder, National Science Foundation (ECCS 2048202) and National Institute of Biomedical Imaging and Bioengineering (REB029541A). K.J., S. Z. and Z.X. acknowledge the support by the National Key R&D Program of China (2019YFA0705000, 2017YFA0303700), Key R&D Program of Guangdong Province (2018B030329001), Leading-edge technology Program of Jiangsu Natural Science Foundation (BK20192001) and National Natural Science Foundation of China (51890861, 11690031, 11621091, 11627810, 11674169, 91950206).

## Author contributions
M.N. and S.W.H. conceived the idea of the experiment. M.N. designed and performed the experiment. Y.X. performed GRIN-MMF modeling and M.N. performed other numerical simulations. K.J., S. Z. and Z.X. fabricated the devices. M.N. and S.W.H. conducted the data analysis and wrote the manuscript. S.W.H. led and supervised the project. All authors contributed to the discussion and revision of the manuscript.

## Competing interests
The authors declare no competing interests.